\documentclass[10pt,journal]{IEEEtran}
\usepackage[T1]{fontenc}
\usepackage[utf8]{inputenc}
\usepackage{textcomp}
\usepackage{microtype}
\usepackage{float}
\usepackage{lipsum}
\usepackage[bottom]{footmisc}
\ifCLASSOPTIONcompsoc
\usepackage[nocompress]{cite}
\else
\usepackage{cite}
\fi

\usepackage{graphicx}
\usepackage{pgfplots}
\usepackage[version=3]{mhchem}
\usepackage{adjustbox}
\ifCLASSOPTIONcompsoc
\usepackage[caption=false,font=footnotesize,labelfont=sf,textfont=sf,subrefformat=parens,labelformat=parens]{subfig}
\else   
\usepackage[caption=false,font=footnotesize,subrefformat=parens,labelformat=parens]{subfig}
\fi

\usepackage{amsmath,amssymb}

\usepackage{multirow,booktabs}% tables
\usepackage{xcolor}

\usepackage{indentfirst}
\usepackage{engtlc}% for symbols including \diff

\usepackage[authormarkup=superscript,deletedmarkup=sout,addedmarkup=em]{changes}% here adding the option "final" would automatically generate the pdf without markup:
\usepackage{soul}

\soulregister\cite7
\soulregister\ref7
\soulregister\pageref7

\usepackage{tikz}

\definecolor{mapblue}{HTML}{0000FF}%
\definecolor{maproyalblue}{HTML}{0088FF}%
\definecolor{mapskyblue}{HTML}{00CCFF}%
\definecolor{mapgreen}{HTML}{00CC00}%
\definecolor{mapdandelion}{HTML}{FFCC00}%
\definecolor{maporange}{HTML}{FF8800}%
\definecolor{mapred}{HTML}{FF0000}%

\usepackage[hyphens]{url}
\usepackage[hidelinks=true,bookmarks=false]{hyperref}
\hypersetup{
  colorlinks   = true, %Colours links instead of colored boxes
  urlcolor     = black, %Colour for external hyperlinks
  linkcolor    = black, %Colour of internal links
  citecolor   = black %Colour of citations
}

\pgfplotsset{compat=1.14} 
\begin{document}
\begin{NoHyper}

\title{Prior Knowledge Input to Improve LSTM Auto-encoder-based Characterization of Vehicular Sensing Data}

\author{Nima~Taherifard, Murat Simsek~\IEEEmembership{Senior~Member,~IEEE}, Charles Lascelles, and Burak~Kantarci,~\IEEEmembership{Senior~Member,~IEEE}

\thanks{N. Taherifard, M. Simsek and B. Kantarci are with the School of Electrical Engineering and Computer Science, University of Ottawa, Ottawa, ON, Canada. E-mails: \{ntahe062, murat.simsek, burak.kantarci\}@uottawa.ca.}
\thanks{Charles Lascelles is with Raven Connected, 441 MacLaren St, Ottawa, ON, K2P 2H3, Canada. Email: charles@ravenconnected.com }
% <-this % stops a space
\vspace{-0.2in}
}

\maketitle
\pagestyle{empty}
\thispagestyle{empty}

\begin{abstract}
Precision in event characterization in connected vehicles has become increasingly important with the responsive connectivity that is available to the modern vehicles. Event characterization via vehicular sensors are utilized in safety and autonomous driving applications in vehicles. While characterization systems have been shown to be capable of predicting the risky driving patterns, precision of such systems still remains an open issue. The major issues against the driving event characterization systems need to be addressed in connected vehicle settings, which are the heavy imbalance and the event infrequency of the driving data and the existence of the time-series detection systems that are optimized for vehicular settings. To overcome the problems, we introduce the application of the prior-knowledge input method to the characterization systems. Furthermore, we propose a recurrent-based denoising auto-encoder network to populate the existing data for a more robust training process. The results of the conducted experiments show that the introduction of knowledge-based modelling enables the existing systems to reach significantly higher accuracy and F1-score levels. Ultimately, the combination of the two methods enables the proposed model to attain 14.7\% accuracy boost over the baseline by achieving an accuracy of 0.96.
\end{abstract}


\IEEEpeerreviewmaketitle

\section{Introduction}

Leveraging precise measuring equipment along with the increasing computing power of the vehicles, the driving event characterization (DEC) systems  are the key components to the road safety of the vehicles in an intelligent transportation system (ITS) where highly reliable connectivity is available between the vehicles \cite{silva.16}. Moreover, the response time improvement introduced by 5G \cite{campolo.17} further incentivizes the investigation and refinement of the DEC for vehicular safety systems. 

Typically, DECs are mainly deployed in distributed sensing platforms such as mobile devices and vehicles on the road and the inertial sensor data is the main instrument used by the systems. Utilizing the in-vehicle inertial sensor data allows for more prompt detection with less lag that can be utilized in the safety applications since inertial data are direct measures of the physical forces applied to the vehicles \cite{ryu.04}.

It is proven that specific driving characteristics like sudden lane changes, acceleration and such are tightly associated as risky driving behaviors \cite{scanlon.17}. To detect these behaviors, the DEC systems follow the objective of detecting anomaly behavior or classifying the pre-defined behavior in the data \cite{taherifard.20.att}. These events include risky driving patterns such as aggressive lane changing, aggressive acceleration, harsh brakes, etc. which occur in small time windows during regular driving patterns and cause the issue of data imbalance. 

The most recent developments in artificial intelligence and machine learning have made the real-time detection of the driving events more feasible and accurate \cite{che.18} by catering accurate methods for analysis of time-series data. More specifically, the recurrent networks which are able to store a brief history of the past input data at any time that makes them the feasible tools for time-series data analysis. Recurrent models such as Long short-term memory (LSTM) architecture allows the machine-learning systems to reserve and consider longer history of the data in the decision making by the networks. Furthermore, storing recent data information is critical for pattern recognition methods employed for driving event characterization systems. LSTM networks are a refined type of recurrent neural networks which are utilized to extract temporal features in time-series data \cite{hochreiter.97}. Such networks are the main enabling factors of the contemporary weather forecasting and language models which heavily depend on the sequential history of the most recent data \cite{sagheer.19}.

\begin{figure}[t]
        \centering
        \includegraphics[width = 0.45\textwidth, trim=0cm 2cm 0cm 0cm, clip]{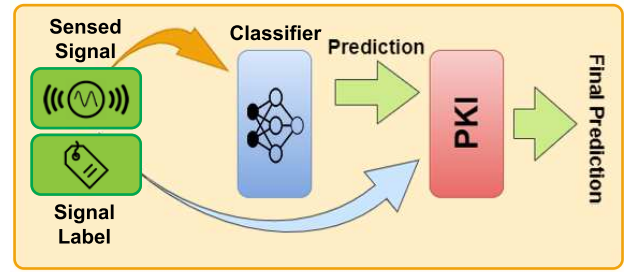}
        \caption{The PKI modulation as module to the classification networks. It accepts the knowledge of the input signals as additional feature for an optimized training process.}
        \label{fig:module}
\end{figure}

In this paper, in order to study the effect of knowledge-based modelling, a previously studied and optimized convolutional LSTM encoder network is chose for baseline experiments. The network allows the system to extract both spatial and temporal features of the signals in order to achieve best performance. Convolutional layers are utilized as spatial feature extraction module since the events have distinct patterns on each axis of the input data while LSTM network performs temporal feature extraction to gain knowledge of the time dependant patterns of the signals. Moreover, a recurrent encoder-decoder network is proposed to not only denoise the input signals but also learn the behavior pattern of individual signals for precise reconstruction of synthetic signals. 

The main contribution of this paper is the introduction of the Prior Knowledge Input (PKI) modelling into machine learning-based recurrent event characterization models.
Knowledge based modelling has been developed to integrate the existing knowledge to the learning process so further improvement can be possible through the mapping between existing knowledge and desired responses. Prior Knowledge Input (PKI) that is one of the knowledge based methods is considered to obtain  the better response  (up to 0.96 accuracy) when compared to the previous work \cite{taherifard.20.att}. PKI utilizes the existing knowledge of driving events as supplementary input alongside with the sensed input signals. As illustrated in Fig. \ref{fig:module}, the PKI is modulated to the existing characterization networks as add-on, which can enable the models to reach the optimal training conditions.

The rest of the paper is organized as follows. Related work is presented in Section II whereas Section III presents the methodology. Performance evaluation and numerical outcome are reported and discussed in Section IV. Finally, Section V concludes the work and offers future intentions.

\section{Related Work}
Numerous studies have been proposed to accelerate the feasibility of the driving event characterization models. Additionally, neural networks have surpassed the machine learning and other intelligent methods in terms of performance. The signal processing domain has also been affected by the aforementioned fields, therefore many researchers are focusing on deep learning methods in order to improve the state-of-the-art event characterization systems. In this section, we first summarize the most recent driving event characterization models and then review the developments of the knowledge-based systems.

\subsection{Driving Event Detection Systems}

Since vehicular sensory data has become increasingly available, the use of neural networks which are highly dependant on big data has become feasible for vehicular applications, including the driving event characterization applications. The systems that utilize such networks are typically focused on visual or sensory data inputs from the vehicles. Ultimately, these models are trained to extract the desired features from the data and identify the anomalies or the pre-defined behaviors from the features. 

A series of studies were conducted on the feature extraction methods to gain knowledge from the inertial or global positional sensed data. Sun et al. \cite{sun.15} propose a system to identify irregular driving on the highways based on accurate positional satellite measurements. A unique calculation method is studied in \cite{zeeman.13} to model the velocity, acceleration, and dynamics of the vehicles. A static threshold value is then applied to the calculated metric in order to detect driving behaviors. Additionally, more studies are focused on the combination modelling of the vehicular measured data. In \cite{eboli.16}, the authors have examined the correlation between the vehicle velocity and lateral acceleration. Investigating this correlation, they conclude that the threshold applied to the lateral acceleration of the vehicle in an event characterization model has to be decreased as the vehicle gains speed. Moreover, smartphone reliant driving event characterization systems are also studied in the literature \cite{saiprasert.17} to utilize the cost-efficient in-built sensors of the smartphones. In \cite{chhabra.19}, the authors designed a system to detect sudden changes of acceleration and unsafe vehicle turns in order to categorise the drivers as aggressive or non-aggressive drivers.

\begin{figure}[t]
        \centering
        \includegraphics[width = 0.50\textwidth, trim=7cm 0cm 0cm 0cm, clip]{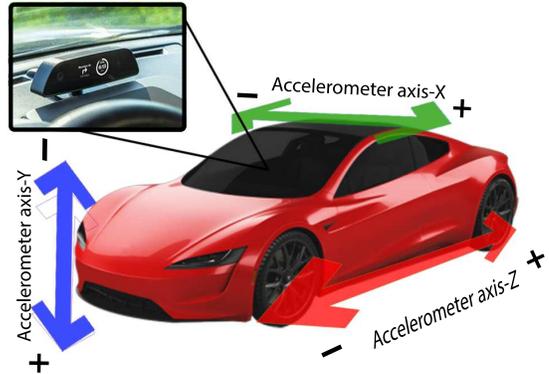}
        \caption{Data collection method. The utilized sensor records acceleration along the three axes of the vehicles.}
        \label{fig:axes}
\end{figure}

Modern day transportation industry is adopting machine-learning systems that can improve through data-driven approaches \cite{soyturk.16,BOUKERCHE.2020101248}. Optimized image processing methods are utilized in \cite{taherifard.19} to categorize driving incidents and make decision on alerting the first-responders to the location. A pre-trained convolutional neural network is used in \cite{he.19} on the phase-space reconstructed vehicle trajectories to evaluate driving behaviors. The study quantitatively evaluates the abnormal behaviors to then detect the abnormal drivers. Moreover, the distracting behaviors that lead to unsafe driving are the subject of various studies. Rao et al. \cite{rao.19} takes the approach of processing the captured camera data via convolutional neural networks and principal component analysis (PCA) to whiten the camera feed and classify the whitened images as distracted and non-distracted driver activities. 

Recurrent neural networks have been improved and made practical more recently. In addition to the progress of various types of recurrent functions, numerous recurrent-based network architectures have been introduced for innovative applications. There has been novel solutions to the major issue of recurrent neural networks, i.e., the vanishing gradient problem which led the networks to face issues in the training process. Long Short-term Memory (LSTM) and Gated Recurrent Units (GRUs) models are among most recent solutions to the issue. Furthermore, there has been numerous studies in order to expand the applications of the recurrent networks. \cite{toderici.17} proposes data compression method utilizing recurrent networks. Combining generative and discriminative networks, a denoising auto-encoder is introduced in \cite{ashfahani.20} in which the model has the ability to take the input features on the fly. An auto-encoder network with the support of random sampling from the encoder latent space is proposed with generative objective in \cite{heljakka.18}. The model mixes a combination of adversarial and reconstruction losses, but unlike Generative Adversarial Networks (GAN) discriminators, the authors employed progressive growing of generator and encoder networks.

\subsection{Knowledge-based Modeling}
%PKI
Knowledge-based modelling is suitable for the driving event classification systems since it only requires the input-output relations \cite{Simsek2016Efficient}. In a conventional neural network, a large training dataset is required to meet the optimal stopping conditions which is not convenient in vehicular applications where there is imbalance to the data. In \cite{simsek.20}, machine-learning detection models are boosted by novel knowledge-based methods and feature selection mechanisms. Utilizing Prior Knowledge Input (PKI) and Prior Knowledge Input with Difference (PKID), the authors are able to significantly improve the average accuracy of the existing machine-leaning methods. The PKI is widely adopted in neural network-based modeling systems such as adversarial task detection in mobile crowd-sensing to microwave modelling for computer aided design \cite{Simsek2011Therecent}. Using the PKI modelling, it is possible to remarkably improve the driving event characterization model in terms of accuracy and overall performance. Indeed, further experimental results are needed which are presented in the following sections.

\section{Methodology}

In this section, the prior knowledge-based signal classification model which is modulated on top of our previous signal generation network is presented in detail. Machine-learning is modelled to imitate the learning process of the brain, and its classification accuracy and the training process efficiency can be improved by modelling the networks utilizing a knowledge-based modulation. We employ a prior knowledge input method to boost our classification performance which is presented in Section \ref{method:C}. A brief introduction of our dataset collection process and the details of our signal generation model are revisited in Section \ref{method:A} and Section \ref{method:B}, respectively whereas the parameters are presented in Table \ref{table:parameters}.

\begin{table}[!t]
\caption{Notation description for mathematical formulas.}

\centering
\begin{tabular}{c|c}
\hline
\multicolumn{1}{l|}{\textbf{Notation}} & \textbf{Description}                 \\ \hline
$x$                                      & Signal features                      \\ \hline
$\hat{x}$                              & Synthetic signal features            \\ \hline
j                                      & Signal dimension                     \\ \hline
i                                      & Signal length                        \\ \hline
C                                      & Number of classes                    \\ \hline
t                                      & Target category of the signal        \\ \hline
$e_{PKI}$                                & Error value of PKI model             \\ \hline
n                                      & Training iteration                   \\ \hline
$P_{c}$                                & Prediction of the classifier network \\ \hline
\end{tabular}
\label{table:parameters}
\end{table}

\subsection{Driving Event Data Gathering}
\label{method:A}

%Data generation method

Both the signal generation network as well as the knowledge-based classification network utilize raw accelerometer sensor data over varied recording sessions. In order to collect the original signals, Raven OBD-II sensor kit \footnote{Raven OBD-II Sensor Kit https://ravenconnected.com/} is mounted on the dashboard of the vehicles. The sensor orientation is perpendicular to the latitudinal, altitudinal, and longitudinal axes of the vehicle as illustrated in Fig. \ref{fig:axes}. 

The collected raw signal data are distributed among five pre-defined risky driving behaviors and on multiple vehicles with different physical features. The pre-defined behaviors include: Aggressive Acceleration (AA), Harsh Braking (HB), Harsh Left Lane Changes (HL), Harsh Right Lane Changes (HR) events in addition to Regular Driving (RD) events.

\begin{figure}[!b]
        \centering
        \includegraphics[width = 0.5\textwidth, trim=0.5cm 0cm 0cm 0cm, clip]{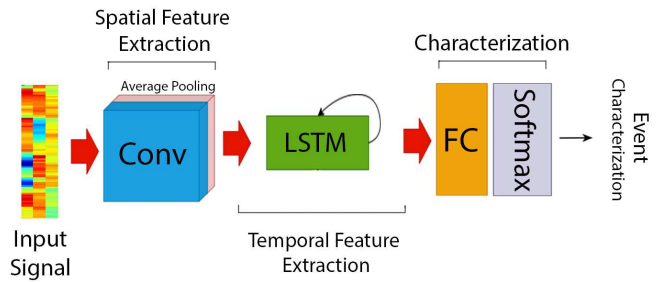}
        \caption{Illustration of the baseline convolutional recurrent neural network. Multiple convolutional and LSTM feature extraction units are utilized before the classification network.}
        \label{fig:convlstm}
\end{figure}

\subsection{Revisiting Synthetic Signal Generation}
\label{method:B}
%Auto-encoder network
In an attempt to overcome the data imbalance and infrequency, we previously proposed an auto-encoder model \cite{taherifard.2020.camad} which is able to accurately extract the underlying behavior of the signals and generate precise synthetic signals. The model consists of two modules, namely, a LSTM recurrent encoder-decoder network which is leveraged in order to populate the training dataset. The second module is used for classification of the signals and is trained on the synthetic dataset.

In order to generate the signals, a windowing mechanism is ran over each individual raw signal. Using the windowing mechanism, the raw signals are split into several signal windows of fixed length with overlapping sections. The overlapping is performed in order to avoid discontinuities in the windowed signals. 

The auto-encoder consists of an encoding network which encodes the input signals to vectors of lower dimension that are feature-rich and contain less noise \cite{yu.13}. Moreover, the encoded vectors are used by the decoding network to reconstruct synthetic signals. The synthetic signal reconstruction is carried out to populate the events in the dataset and to strengthen the classification training process by mitigating data imbalance. 

To train the recurrent auto-encoder for synthetic signal generation, the signal windows are infused with randomly generated noise and fed as the input to the auto-encoder network. The recurrent auto-encoder network concludes a two-layer LSTM network as the encoder which maps the signal windows into fixed-size vectors. It is worth mentioning that the conversion of the signals into lower dimension vectors by the encoder network has the property of signal denoisification since high dimensional data is often more convenient to learn in lower dimensions as it contains less noise. The decoder network attempts to reconstruct the original signals from the vectors. We set the network to generate 10 times synthetic signal windows for the training process of the classification module. In the training process, the auto-encoder network is trained in self-supervised manner with the intention of minimizing the mean absolute error (\ref{mae}) of the original and generated signals. 

\begin{equation}
    \sum_j \sum_i \left \| x^{\left ( j \right )} - {\hat{x}}^{\left (j  \right )} \right \|^{2} 
    \label{mae}
\end{equation}

In order to demonstrate the improvement introduced by the PKI based modelling proposed in this work, we trained a feed forward multi-layer perceptron (MLP) of 3 layers and a convolutional recurrent classification (ConvLSTM) model which adds the spatio-temporal feature extraction layers to the MLP network. The ConvLSTM model utilizes multiple feature extraction layers of decreasing sizes as illustrated in Fig. \ref{fig:convlstm}. However, the objective of both classification networks is to categorize the signals into 5 pre-defined driving event categories using categorical cross-entropy which is implemented as the softmax (\ref{sm}) of cross-entropy function (\ref{cce}).

\begin{equation}
    f\left ( x \right )=\frac{e^{x}}{\sum_C e^{x}} 
    \label{sm}
\end{equation}

\begin{equation}
    CE = -\sum_{C} t \; log(f(x)) 
    \label{cce}
\end{equation}

\subsection{Prior Knowledge Input}
\label{method:C}

To boost the performance of the classification networks on the driving event signals, we propose knowledge-based schemes. The Prior Knowledge Input (PKI) method, specifically, is used to incorporate the existing knowledge of the signals into the characterization process. Doing so allows for a more efficient and less complex training process which requires less amount of training data \cite{simsek.20} and aids the characterization systems to push past the limited accuracy of deep learning methods and achieve the potential accuracy rates \cite{Simsek2016Efficient}. Moreover, any modelling scheme can be utilized in order to generate the input knowledge when there is no available prior knowledge of the inputs. Though, neural networks are often the method of choice for the knowledge-based modelling \cite{simsek.17}.

The PKI model embeds the experience/knowledge of the signal category into training process which allows for reduction of model complexity through augmentation of the model inputs. The training and testing process of the PKI model is shown in Fig \ref{fig:pki}. The PKI network is implemented in 2 hidden layers that utilize $tanh$ activation function with a softmax classification layer to output the event prediction. The training process can be formulated with (\ref{pki_train}) while iterating to lower the margin of the network predictions and the prior knowledge (\ref{pki_error}) for $n$ training iterations. The inference process is also calculated by (\ref{pki_infer}) as illustrated in Fig. \ref{fig:pki}.

\begin{equation}
    \left ( PKI \right )_{Train} = \arg\min_{x} {\left \|\; ... \;\;\; e^{(n)^T}_{PKI} \;\;...\; \right \|}
    \label{pki_train}
\end{equation}

\begin{equation}
    e^{(i)}_{PKI} = C_{events} (x^{(i)}) - PKI (x^{(i)} , P^{(i)}_{c})
    \label{pki_error}
\end{equation}

\begin{equation}
    P_{PKI} = (PKI)_{Train} (x_ , P_{c})
    \label{pki_infer}
\end{equation}

\begin{figure}[!t]
        \centering
        \includegraphics[width = 0.45\textwidth, trim=0cm 0cm 0cm 0cm, clip]{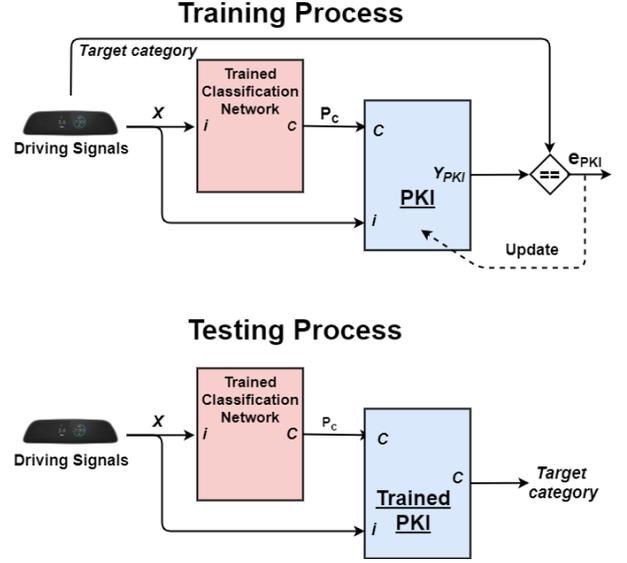}
        \caption{Training and testing process of the prior knowledge input method to improve existing performance of driving event characterization systems.}
        \label{fig:pki}
\end{figure}

\section{Experimental Evaluation}

\subsection{Experimental Settings}
\label{eval.settings}
All the experiments are performed on the gathered dataset of 70 driving event sessions populated by 10 times synthetic signals using the auto-encoder network. To keep the experiments credible, all tests are performed on an identical testset.

Initially, accelerometer sensor signals with variable duration of 2000 to 3600 milliseconds are captured. The accelerometer recording frequency is set to 25 Hz. In total, 70 driving event sessions are recorded. The distribution of the collected dataset is demonstrated in Table \ref{table:data}. Originally, 12 AA, 13 HB, 16 HL, 15 HR, and 14 RD driving sessions were accumulated in the dataset. The sessions are split into 572 fixed-size event windows through the sliding mechanism using 600 millisecond window length and 50\% overlap settings as demonstrated in Fig. \ref{fig:window}. Last but foremost, a randomly selected training dataset of 70\% (400) sliced windows is selected and 30\% (172) sliced windows are kept unseen by the systems for testing.

\begin{table}[!t]
\caption{Training signal count distribution before and after slicing over 5 classes.}
% \vspace{0.6}
\centering
\begin{tabular}{c|c|c}
\hline
\multicolumn{1}{l|}{\textbf{Event Type}} & \multicolumn{1}{l|}{\textbf{Event Sessions}} & \multicolumn{1}{l}{\textbf{Sliced Windows}} \\ \hline
HB                                   & 13                                         & 104                                       \\ \hline
AA                                   & 12                                         & 108                                       \\ \hline
HL                                   & 16                                         & 126                                       \\ \hline
HR                                     & 15                                         & 121                                       \\ \hline
RD                                   & 14                                         & 113                                       \\ \hline
\end{tabular}
\label{table:data}
\end{table}

\begin{figure}[!t]
        \centering
        \includegraphics[width=0.75\linewidth, trim=0cm 0cm 0cm 0cm, clip]{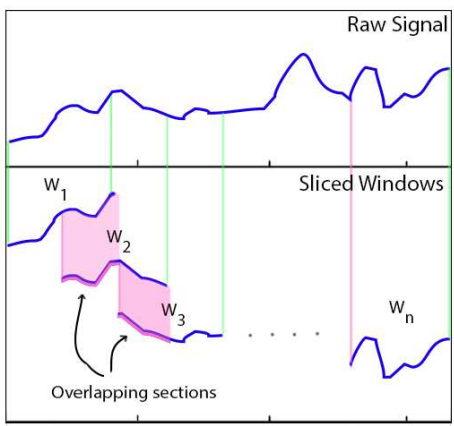}
        \caption{Abstract visualization of the signal slicing performed before the training process. The signals are sliced into 6000 ms windows with 0.5 overlap value.}
        \label{fig:window}
\end{figure}

\begin{figure}[!t]
        \centering
        \includegraphics[width = 0.65\linewidth, trim=0cm 0cm 0cm 0cm, clip]{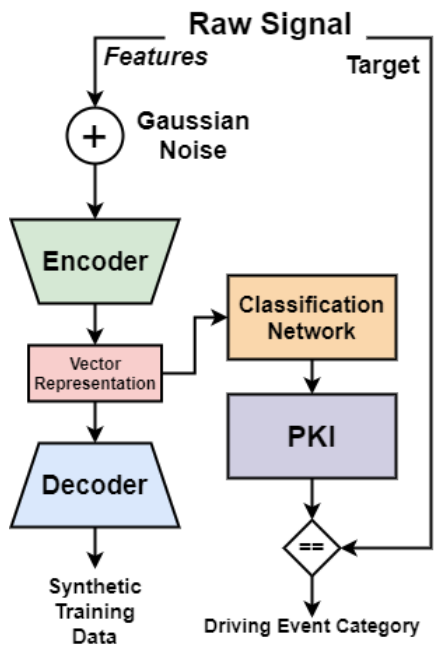}
        \caption{Training process flow of the PKI method. The vector representation of the noisy raw signals are extracted from the encoder network. The vectors are utilized by the decoder network for synthetic signal generation and by the PKI classification network for characterization of the events.}
        \label{fig:system}
\end{figure}

The synthetic data generation is don on the training data through the auto-encoder network with 3 LSTM network with decreasing hidden layers, from 1000 to 300 cells. However, each individual input signal is first duplicated into 10 individual signals and each is infused with random Gaussian noise. The noisy signals are mapped to vectors of size 300 at the final hidden layer of the encoder network. Subsequently, the decoder network is implemented as a mirrored LSTM network with the same dimensions to the encoder and attempts to reproduce the raw signals from the noisy inputs. Furthermore, the decoder output (i.e. synthetic training data) is used in the training process of the classifier networks. However, in the inference process, the decoder network is deactivated and the encoded vectors are directly passed to the classifier network. The process flow of the system is depicted in Fig. \ref{fig:system}.

In order to demonstrate the impact of the PKI module, we chose a fully connected feed forward classifier network, more specifically a Multi-Layer Perceptron (MLP), as baseline as well as an optimized convolutional recurrent classifier network (ConvLSTM) for further experimentation. Moreover, the classification modules perform a supervised task utilizing categorical cross-entropy for optimization on each training iteration and aim to lower the margin of the predicted events and the target categories. Adam optimizer with training parameter of $\beta_1 = 0.9$ and $\beta_1 = 0.95$ and $\epsilon = 10^{-8}$ is selected for both classifier networks. Lastly, the networks are trained with learning rate of 0.03 until the stopping is triggered.

\subsection{Numerical Results}
\label{eval.results}

As mentioned earlier, the MLP and ConvLSTM classification networks are trained on our driving event dataset with and without the PKI modulation. Additionally, the networks are trained and tested on raw signals, in the absence of the reconstructed synthetic data, for a more comprehensive demonstration of the PKI benefits. Through testing of the models on identical testset, the impact of the PKI model is demonstrated in Table \ref{table:results}.

Collecting the experimental results, the performance improvement of PKI modulation can be observed across all the tested classification models. The experiments on each classifier network are performed over 10 separate runs for statistical presentation of the results which are stated with confidence levels of 95\%. While the baseline MLP classifier gained the most significant accuracy improvement (by average of over 5\%), the PKI modulation proved to increase our most accurate model (i.e. auto-encoder with ConvLSTM classifier) to over 96\% average accuracy as shown in Fig. \ref{fig:bars}. Last but not least, as a benefit of PKI accuracy improvements, the decrease in false positive cases of the models is reflected in improved precision, recall, and therefore F1-score values across the models.

\begin{figure}[!t]
        \centering
        \includegraphics[width=0.99\linewidth, trim=0cm 0cm 0cm 0cm, clip]{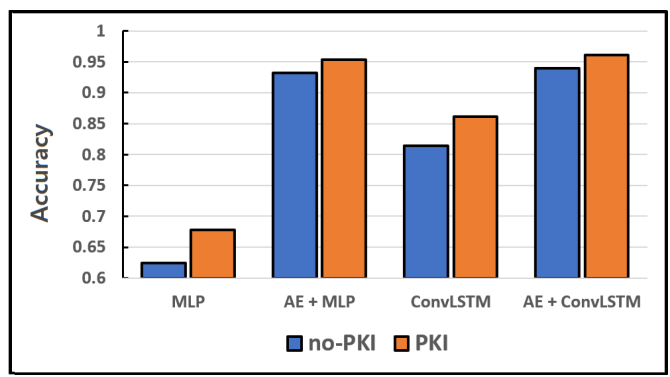}
        \caption{Prior Knowledge Input modulation impact comparison.}
        \label{fig:bars}
\end{figure}

\begin{table*}[!t]
\caption{Numerical results.}
\centering
\begin{tabular}{c|c|c|c|c|c}
\hline
\multicolumn{1}{l|}{\textbf{PKI Modulation}} & \textbf{Model} & \textbf{Accuracy}                           & \textbf{Precision}          & \textbf{Recall}                                   & \textbf{F1-Score}           \\ \hline
OFF                                          & MLP           & 0.624$\pm$0.014                                           & 0.27$\pm$0.011                           & 0.30$\pm$0.012                                                 & 0.29$\pm$0.012                           \\ \hline
OFF                                          & ConvLSTM        & 0.814$\pm$0.009                 & 0.77$\pm$0.010 & 0.79$\pm$0.009                       & 0.78$\pm$0.009 \\ \hline
OFF                                          & AE + MLP       & 0.932$\pm$0.007 & 0.89$\pm$0.008 & 0.90$\pm$0.008                       & 0.89$\pm$0.007 \\ \hline
OFF                                          & AE + ConvLSTM  & 0.940$\pm$0.004                 & 0.12$\pm$0.003 & 0.11$\pm$0.002                       & 0.11$\pm$0.003 \\ \hline \hline
ON                                           & MLP            & 0.678$\pm$0.009                 & 0.39$\pm$0.010 & 0.44$\pm$0.009                       & 0.42$\pm$0.009 \\ \hline
ON                                           & ConvLSTM       & 0.861$\pm$0.008                                           & 0.68$\pm$0.008                           & 0.62$\pm$0.010                                                 & 0.65$\pm$0.009                         \\ \hline
ON                                           & AE + MLP       & 0.954$\pm$0.006                 & 0.87$\pm$0.008 & 0.90$\pm$0.006 & 0.88$\pm$0.007 \\ \hline
ON                                           & AE + ConvLSTM  & 0.961$\pm$0.003                 & 0.87$\pm$0.002 & 0.93$\pm$0.002                       & 0.90$\pm$0.002 \\ \hline
\end{tabular}
\label{table:results}
\end{table*}

\section{Conclusion}
This paper has presented a novel approach to boost the classification performance for risky driving events that are recognized from vehicular sensors. To do so, the integration of a Prior Knowledge Input (PKI) modelling into the event characterization networks has been proposed to not only improve the overall classification accuracy but also reduce the false detections. The PKI model leverages the existing knowledge of the input signals as an supporting feature in order to lower the complexity of the model and therefore improve the detection accuracy of the baseline classification networks. The results prove that the performance of the classification models can be significantly improved with the introduction of the PKI modelling. Our results reveal that using the PKI method improves the performance of the baseline MLP classifier by over 5\%. Similarly the ConvLSTM network experiences a 4.7\% improvement when coupled with the prior knowledge module while the performance of the auto-encoder with ConvLSTM is also improved by an additional 2\%.

Future extensions of this work include extending the experiments to run on larger datasets where data scarcity is not a challenge. Furthermore, characterization of additional behaviour classes is also included our ongoing work.

\section*{Acknowledgment}

This work was supported in part by the Natural Sciences and Engineering Research Council of Canada (NSERC) DISCOVERY RGPIN/2017-04032 and Ontario Centres of Excellence (OCE) TalentEdge Internship Project 32815 in collaboration with Raven Connected.

\bibliographystyle{ieeetr}
\bibliography{refs.bib}
\pagestyle{empty}
\thispagestyle{empty}

\end{NoHyper}
\end{document}